\documentclass[conference]{IEEEtran}
\IEEEoverridecommandlockouts
\usepackage{cite}
\usepackage{amsmath,amssymb,amsfonts}
\usepackage{algorithmic}
\usepackage{graphicx}
\usepackage{textcomp}
\usepackage[usenames,dvipsnames]{xcolor}
\usepackage{svg}
\usepackage{enumitem}  
\usepackage{verbatimbox}
\usepackage{subcaption}
\usepackage{float} 
\usepackage{stfloats}

\usepackage{xcolor}
\usepackage{listings}

 \usepackage{url}

\usepackage[capitalise]{cleveref}
\usepackage{booktabs}
\usepackage{multirow}
\usepackage{makecell}
\usepackage{pifont}

\def\BibTeX{{\rm B\kern-.05em{\sc i\kern-.025em b}\kern-.08em
    T\kern-.1667em\lower.7ex\hbox{E}\kern-.125emX}}

\usepackage{titlesec}

\titleformat{\section}
  {\bfseries\normalsize}{\thesection}{1em}{}
\titleformat{\subsection}
  {\bfseries\normalsize}{\thesubsection}{1em}{}

\begin{document}

\definecolor{darkgreen}{RGB}{9, 107, 104}
\definecolor{darkred}{RGB}{72, 104, 101}
\definecolor{darkyellow}{RGB}{245, 238, 221}
\definecolor{myblue}{RGB}{78, 147, 217}

\newcommand{\varcolor}{darkgreen}
\newcommand{\opcolor}{myblue}
\newcommand{\strcolor}{darkyellow}

\newcommand{\vartext}[1]{\textcolor{\varcolor}{#1}}  
\newcommand{\optext}[1]{\textcolor{\opcolor}{#1}}   
\newcommand{\strtext}[1]{\textcolor{\strcolor}{#1}}

\lstdefinelanguage{QuantumIR}{
  morekeywords={ensemble, eir, gate, arith, remsi, if, index_cast, constant, f64, virtual_qubit, physical_qubit, alloc_cbits, program_alloc, quantum_program_iteration, index, extract, apply, measure, transmit_results, int_uniform, i32, x10xi32, <10, to, step, float_uniform, gate_distribution, apply_distribution, scf, for},
  sensitive=true,
  alsoletter={\%},
  morestring=[b]",
  comment=[l]{//},
  literate={.}{{.}}1
}

\lstdefinestyle{quantumstyle}{
  language=QuantumIR,
  basicstyle=\fontfamily{pcr}\selectfont\small,
  keywordstyle=\color{\opcolor},
  stringstyle=\color{\strcolor},
  identifierstyle=\color{\varcolor},
  commentstyle=\color{gray}\itshape,
  columns=fullflexible,
  keepspaces=true,
  showstringspaces=false,
}

\newcommand{\qinline}[1]{\textbf{\lstinline[style=quantumstyle]!#1!}}

\newcommand{\framework}{ensemble-IR}
\newcommand{\Framework}{Ensemble-IR}

\newcommand{\xmark}{\color{darkred}\ding{55}}%
\newcommand{\cmark}{{\color{darkgreen}\ding{51}}}
\newcommand{\ycmark}{{\color{darkyellow}\ding{51}}}

\newcommand{\compilationtimereduction}{70\%}
\newcommand{\runtimereduction}{30\%}
\newcommand{\filesizereduction}{75\%}

\newcounter{refnum}  
\setcounter{refnum}{1}  
\newcommand{\citeauto}{[\arabic{refnum}]\stepcounter{refnum}}  
\newcommand{\setcitenum}[1]{\setcounter{refnum}{#1}}  

\title{ 
\begin{center}
{\Large \textbf{\Framework{}: Concise Representation for Quantum Ensemble Programs}}
\end{center}
}

\begingroup
\small
\author{
\IEEEauthorblockN{Sourish Wawdhane*}
\IEEEauthorblockA{\textit{The University of Texas, Austin}}
\and
\IEEEauthorblockN{Sashwat Anagolum*}
\IEEEauthorblockA{\textit{SolarWinds}}
\and
\IEEEauthorblockN{Poulami Das}
\IEEEauthorblockA{
\textit{The University of Texas, Austin}
} 
\and
\IEEEauthorblockN{Yunong Shi}
\IEEEauthorblockA{
\textit{AWS Quantum Technologies}
}

}
\endgroup
\IEEEaftertitletext{\vspace{-1.7em}}

\maketitle

\begingroup
\renewcommand\thefootnote{}
\footnotetext{*Both authors contributed equally to this research. All authors can be reached at sourishw@utexas.edu, sashwat.anagolum@solarwinds.com, poulami.das@utexas.edu, and shiyunon@amazon.com respectively.}
\addtocounter{footnote}{-1}
\endgroup

\begin{figure*}[b]
    \centering
    \includegraphics[width=\linewidth]{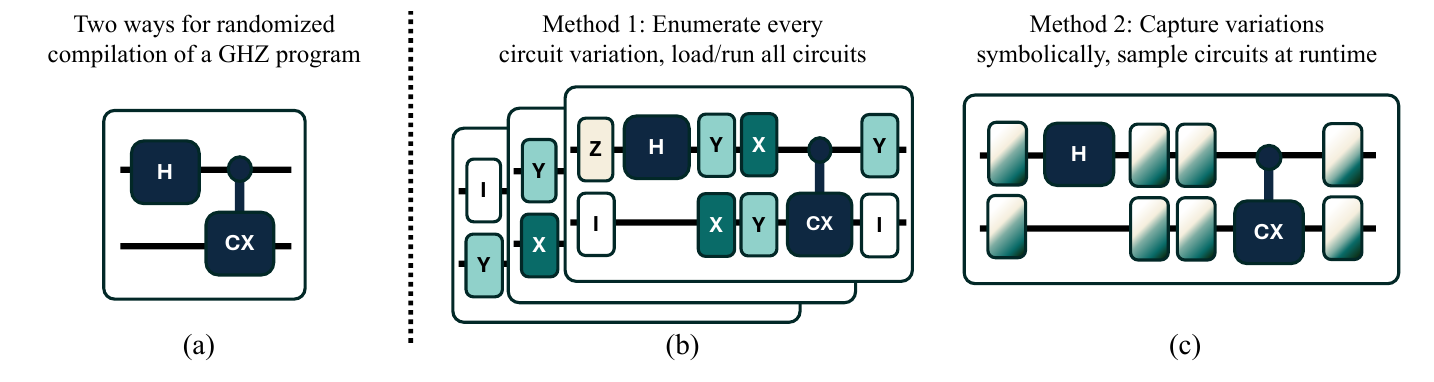}
    \caption{(a) A two-qubit GHZ program to which the randomized compilation error mitigation method is applied. (b) Current frameworks express ensemble workloads by enumerating circuits. (c) \Framework{} expresses a shared program structure and captures the circuit generation framework associated with an ensemble program. The complete ensemble code for this program is included in the appendix.}
    \label{fig:comparative}
\end{figure*}

\section{Overview of Quantum Ensemble Programs}
Multiple important classes of quantum computing algorithms such as error mitigation~\cite{czarnik_clifford_data_regression,tiron_zero_noise_extrapolation,temme_probabilistic_error_cancellation,das_jigsaw,hashim_randomized_compilation}, device characterization~\cite{proctor_rb_via_mirro_circuits,magesan_randomized_benchmarking,flammia_average_circuit_eigenvalue_sampling,cross_quantum_volume,erhard_rb_via_cycle_benchmarking,wack_clops_benchmark}, quantum information protocols~\cite{satzinger_characterizing_topological_order,zhu_shadow_fidelity_estimation,huang_efficient_estimation_of_pauli_observables,levy_shadow_quantum_process_tomography,huang_predicting_many_properties_from_very_few_measurements}, circuit knitting~\cite{lowe_wire_cutting_via_randomized_measurements,schmitt_gate_cutting}, and composite workloads~\cite{huang_predicting_many_properties_from_very_few_measurements,chen_robust_shadow_estimation,majmudar_zne_best_practices,lowe_vncdr} involve running large ensembles of related quantum circuits, often numbering in thousands of unique circuits. 

These multiple circuits share a largely similar structure - connectivity patterns, stages of gates, and overall layout are often preserved across the family. However, individual circuits differ from each other in a few key ways, including changes in circuit width and depth, placement of one and two-qubit gates within layers, changes in gate types and parameter values, and more. Circuits within these programs are often produced by a fixed generation process that sometimes involves randomness.

\section{Limitations of Existing Frameworks}

Leading quantum programming frameworks and IRs such as OpenQASM~\cite{cross2017openquantumassemblylanguage}, Qiskit~\cite{javadiabhari2024quantumcomputingqiskit}, QIR~\cite{luo2023formalizationquantumintermediaterepresentations}, and Mitiq~\cite{LaRose_2022} are not designed for concise representation of quantum ensemble programs. OpenQASM and QIR provide functionality to express hybrid quantum-classical algorithms and parametric gates. However, they do not provide primitives to concisely express variations in gate placement, gate types, qubit operand selection, or circuit structure. Qiskit and Mitiq provide Python-based APIs that let users generate families of related quantum circuits programatically. However, before these circuits are compiled and executed, they are unrolled into seperate circuits. A commonality tying these quantum frameworks is their reliance on fully materialized circuit representations. Today's frameworks serialize and load these ensemble workloads onto quantum systems as sets of unique circuits.  

Figure~\ref{fig:comparative} depicts current methods of expressing ensemble workloads. Applying the randomized compilation \cite{Wallman_2016} error mitigation strategy to a two-qubit GHZ~\cite{ghz_state_paper} program depicted in Figure~\ref{fig:comparative}(a) requires running many individual circuits to infer a final program result. Existing frameworks express this program by enumerating the circuits to be run. Figure~\ref{fig:comparative}(b) depicts the differences between these circuits, involving adding randomly selected pairs of pauli gates before and after each cycle of gates from the original GHZ program. Many ensemble workloads require enumeration of a large number of unique circuits, often ordering in the tens of thousands.





\begin{figure*}[t]
    \centering
    \includegraphics[width=\linewidth]{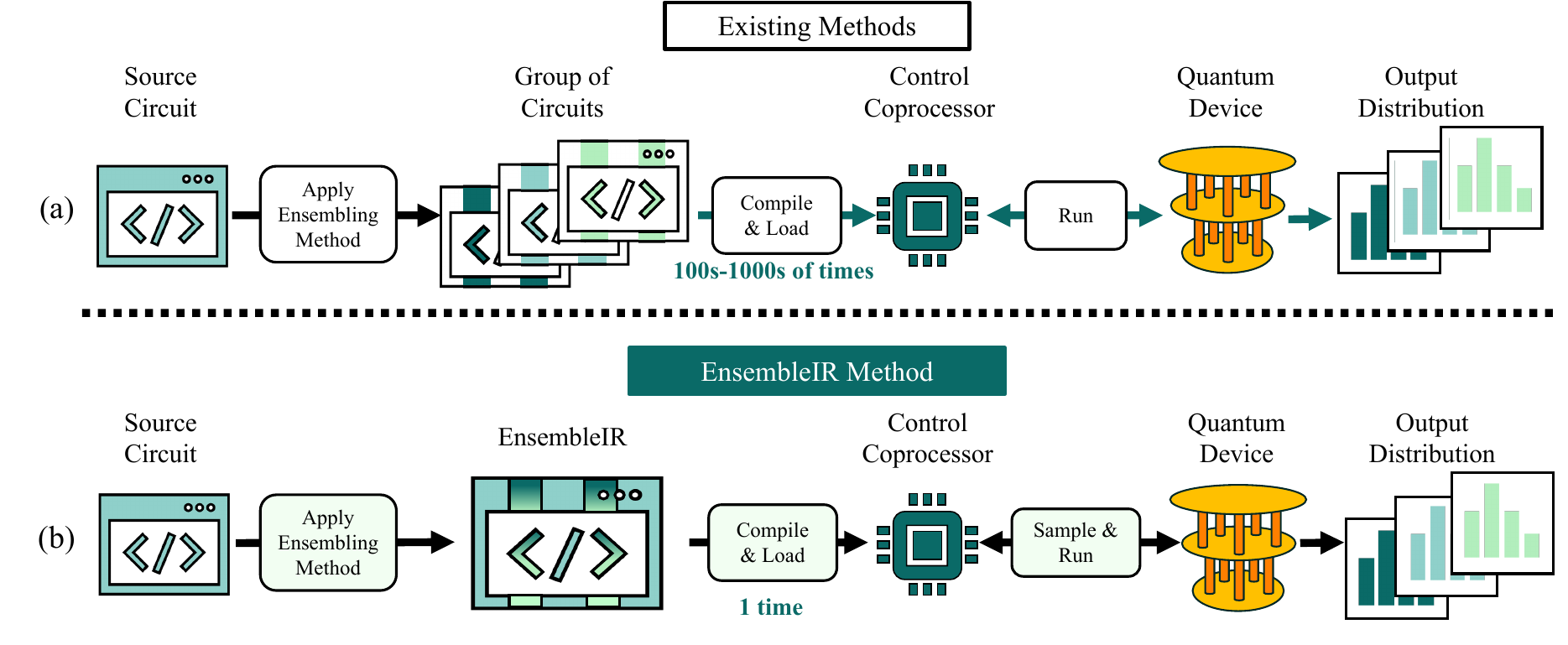}
    \caption{(a) Today's quantum systems run ensemble workloads by compiling and loading each circuit separately onto quantum control coprocessors. (b) \Framework{} enables ensemble workloads to be loaded onto control systems as a single program structure. Contained circuits can be recreated on the fly at runtime.}
    \label{fig:bigpicture}
\end{figure*}
\section{\Framework{}}
In this paper, we propose \Framework{}, an intermediate representation designed to concisely express quantum ensemble programs. Instead of enumerating the circuits within a workload, \Framework{} outlines programs with a single concise circuit generation framework, capturing the similarities and variations between sets of circuits, as illustrated in Figure~\ref{fig:comparative}(c). \Framework{} enables entire workloads to be expressed and loaded as a single program onto quantum control systems, as shown in Figure~\ref{fig:bigpicture}(b). This is in contrast to existing systems, which load groups of circuits onto control hardware separately, as depicted in Figure~\ref{fig:bigpicture}(a). In the following section, we describe a subset of the operations \Framework{} introduces to concisely represent ensemble programs.
\subsection{Overview of \Framework{}}


 \Framework{} expresses circuit generation logic using control flow constructs, distributions across qubits and gates, random number generation primitives, and symbolic parameters. These instructions enable a control-coprocessor to sample individual circuits from a program description at runtime.

\subsection{Qubits and Simple Gates}

\newenvironment{centeredverbatim}{\expandafter\verbatim\centering}{\endverbatim}
We introduce \Framework{} as a dialect within the Multi-Level Intermediate Representation (MLIR) framework. This section describes simple primitives. In the \textbf{\qinline{ensemble}} dialect,  qubits are typically organized into tensors to facilitate indexed addressing. For instance, a register of 25 qubits is represented as a tensor of type 
\qinline{tensor<25x\!eir.physical_qubit>}


The \qinline{ensemble} dialect does not restrict the set of allowed quantum gates, allowing system providers and users to express any gates they choose. Quantum gates may be specified in \qinline{ensemble} using the \qinline{eir.gate} operation. In addition, parameters for each gate may be procedurally determined at runtime. For example:

\begin{figure}[H]
    \centering
    \includegraphics[width=0.8\linewidth]{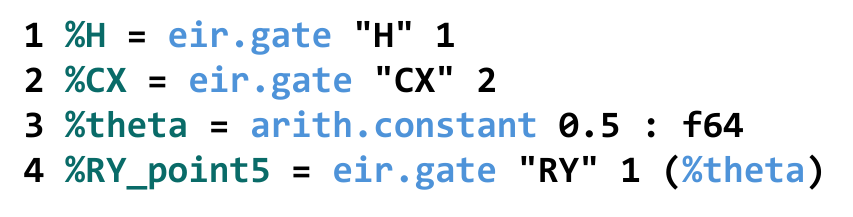}
\end{figure}
constructs a single-qubit Hadamard (H) gate, a two-qubit controlled-NOT (CX) gate, and a single-qubit Y axis rotation (RY) gate with a rotation parameter of 0.5 radians. These gate definitions are then used in conjunction with qubit operands to apply quantum operations within the circuit. A complete example of a two-qubit GHZ circuit is included in the appendix to demonstrate these primitives.

\subsection{Random Number Generation} Many ensemble workloads rely on randomness to introduce variations between their circuit instances. \Framework{} supports these workloads by introducing built-in random number generation primitives. The operations \qinline{eir.int_uniform} and \qinline{eir.float_uniform} allow users to generate tensors of uniformly distributed random numbers within a user-specified range at runtime. These operations can be used to randomize gate choices, parameter values, qubit selections, and more without requiring explicit external precomputation. For example in the following code snippet:


\begin{figure}[h]
    \centering
    \includegraphics[width=1\linewidth]{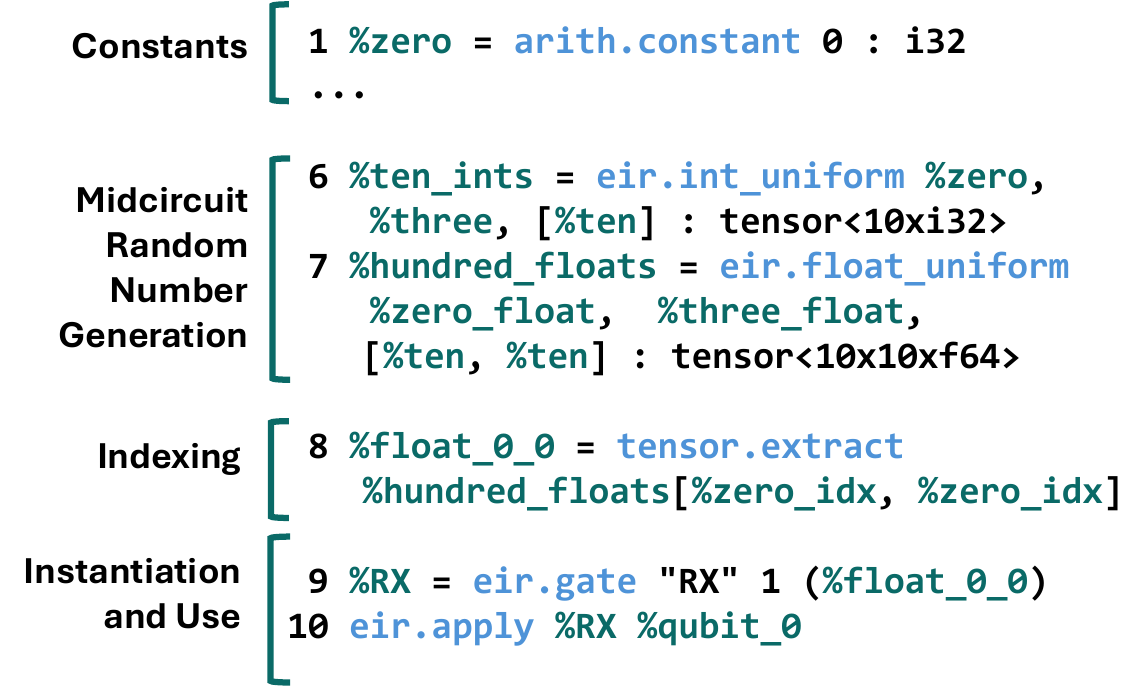}
\end{figure}

\qinline{\%ten_ints} is constructed as a tensor of ten integers each uniformly sampled from the set \qinline{\{0, 1, 2\}}. Similarly, \qinline{\%hundred_floats} is constructed as a  10x10 tensor of floating point numbers, each uniformly sampled from the range \qinline {[0.0, 3.0)} at runtime. On line \qinline{8}, the \qinline{(0, 0)} element of \qinline{\%hundred_floats} is indexed, then used as a parameter to a single-qubit X axis rotation (RX) quantum gate on line \qinline{9}. 

\begin{figure*}[t]
    \centering
    \includegraphics[width=0.8\linewidth]{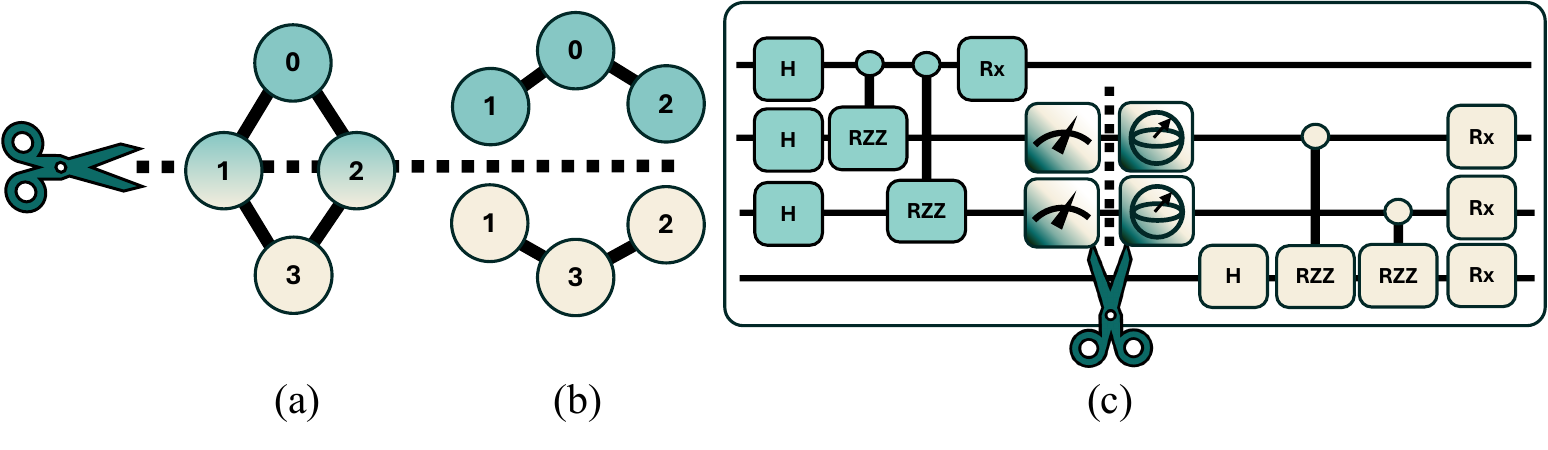}
    \caption{(a) The topology of a 4-qubit QAOA circuit, with wire cutting applied to it to form a (b) 3-qubit top subcircuit and a 3-qubit bottom subcircuit. Multicolored icons represent "ensembled" measurement basis shifts and state preparation operations. (c) The 4-qubit wire-cut circuit depicted with sub circuits illustrated.}
    \label{fig:qaoacut}
\end{figure*}

\newpage

\subsection{Probabilistic Gate Sampling and Qubit Selection}
Ensemble workloads can use random numbers generated at runtime to choose a quantum gate from a predefined set for each circuit iteration. The \qinline{eir.gate_distribution} operation represents a collection of such gate candidates, with all candidates sharing the same number of qubit operands. We illustrate the construction of a gate distribution and its usage in the following code snippet. Here the prior defined \qinline{\%ten_integers} random number tensor is used to randomly select a gate to apply to \qinline{\%qubit_0}:

\begin{figure}[h]
    \centering
    \includegraphics[width=1\linewidth]{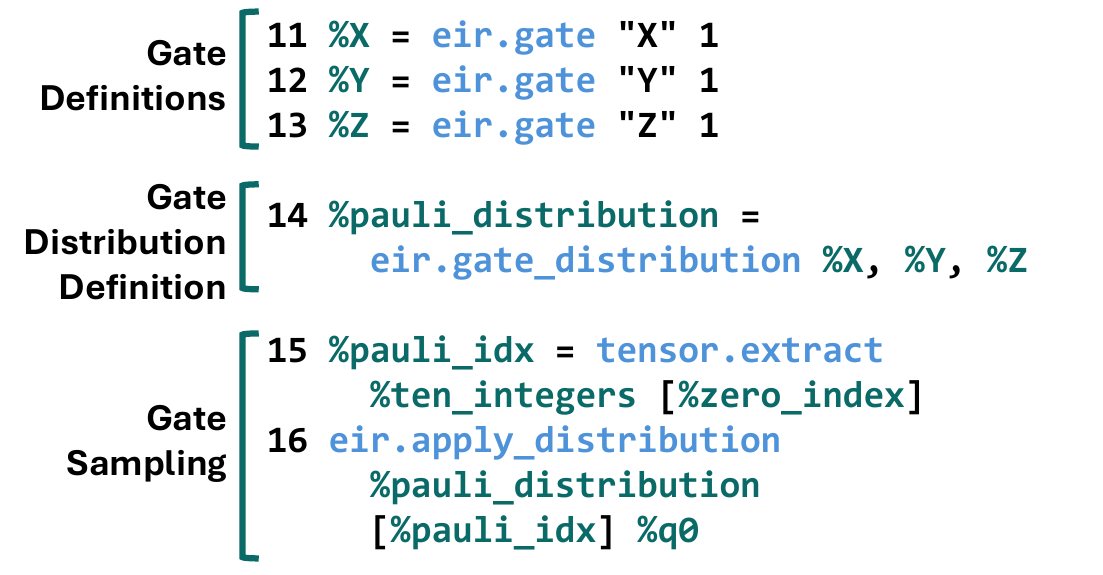}
\end{figure}

\Framework{}'s pattern of dynamic gate selection provides a powerful abstraction for specifying circuit behavior without hardcoding every variant. By defining a gate pool and sampling logic, users can compactly express a wide range of circuit variations, each corresponding to a different realization of the ensemble. In the following section, we use these features to encode the variations contained within a wire-cut program. 

\subsection{Inter-Circuit and Intra-Circuit Control Flow}
Control flow primitives from the \qinline{scf} dialect in MLIR enable fine-grained circuit specialization within and across circuit iterations. We demonstrate circuit specialization using wire cutting \cite{Lowe_2023} as an example. We apply wire cutting to a to a 4-qubit toy Quantum Approximate Optimization Algorithm (QAOA) \cite{farhi2014quantumapproximateoptimizationalgorithm} circuit to form two 3-qubit subcircuits, as depicted in Figure~\ref{fig:qaoacut}. Depending upon the circuit iteration, either the top subcircuit or the bottom subcircuit is executed. When the top subcircuit is executed, gates for measurement basis shifts are applied to each of the qubits using an \qinline{scf.for} loop. If the bottom subcircuit is executed, a series of state preparation operations are applied before application of Hadamard (H) and RZZ entanglement gates in the subcircuit. Both sets of operations are selected at runtime based on a circuit iteration index. 


\begin{figure}[h]
    \centering
    \includegraphics[width=1\linewidth, trim=0 10 0 0, clip]{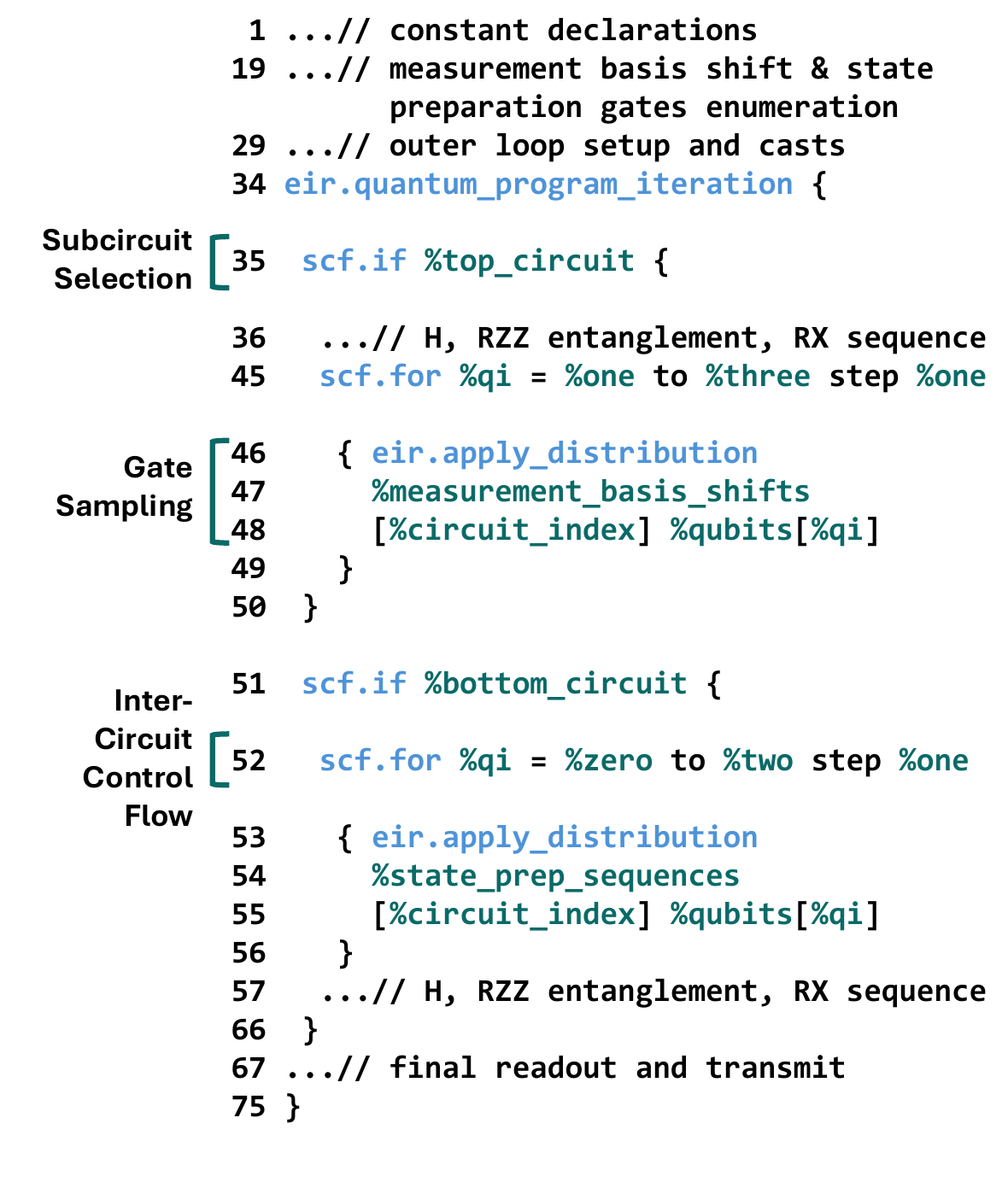}
\end{figure}

The tandem use of control flow operations for subcircuit selection and ensemble gate distributions for basis shifts and state preparation illustrates how users can express different wire-cutting iterations using the same IR. Following the defined control flow, the quantum coprocessor can preprocess the program description into many individual circuits one at a time before running each circuit on quantum hardware.

\newpage

\section{\Framework{} Sample Programs}
To illustrate the full feature set of \Framework{}, we implement a set of 18 sample ensemble programs  based on experiments performed in prior work in domains such as quantum information, quantum characterization, verification, and validation (QCVV), error mitigation, and circuit knitting.
We also include a set of composite workloads requiring combinations of two or more circuit ensembling techniques, also from previous works. Source code, the EnsembleIR MLIR dialect, and details of program parameters are available at \textcolor{blue}{\url{https://github.com/SashwatAnagolum/ensemble-compilation}}.



\section{Conclusion}

\Framework{} introduces an intermediate representation designed explicitly for quantum ensemble programs, enabling concise encoding of structurally related circuits. Our implementation of a variety of different workloads demonstrates that an ensemble-centric IR is practical and essential for scalable quantum software.

\section{Acknowledgments}
Poulami Das acknowledges support through the AMD endowment at UT Austin. The authors thank Erik Davis and Mitch D'Ewart for their feedback and discussion. We thank Avinash Kumar for editorial feedback.

\bibliographystyle{unsrt}
\bibliography{refs}

\section{Appendix}

This appendix provides additional code snippets for workloads introduced in the main text. We present a baseline two-qubit GHZ circuit and its transformed variant using randomized compilation for error mitigation. Together, they highlight how simple circuits can be procedurally transformed into ensembles using \Framework{}'s abstractions.
\subsection{Two-Qubit GHZ Circuit}

Illustrated below is a complete code snippet for a simple two-qubit GHZ circuit, depicted graphically in Figure~\ref{fig:comparative}(a). 

\begin{figure}[h]
    \centering
    \includegraphics[width=1\linewidth]{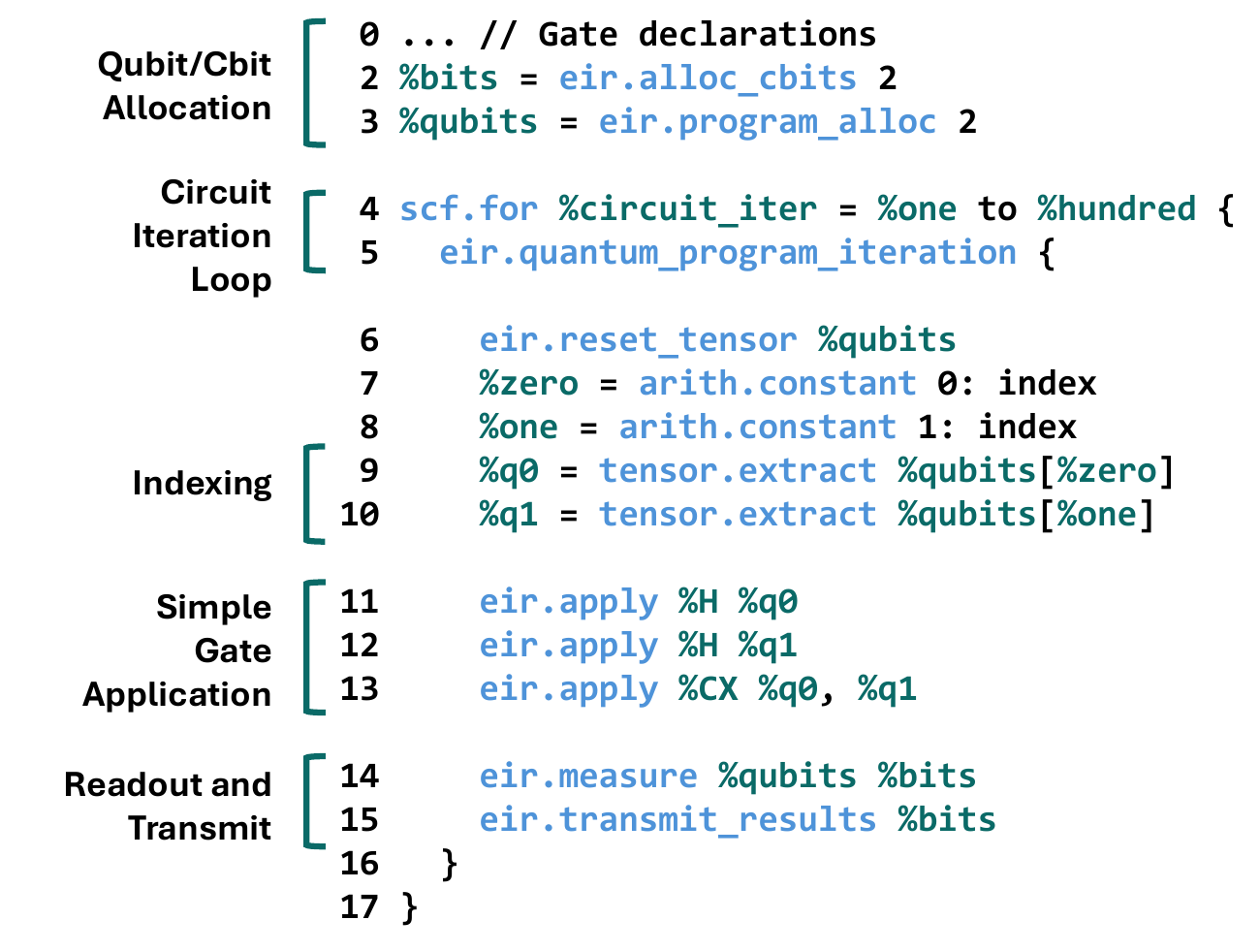}
\end{figure}

In this example, two classical bits and two qubits are defined. The \qinline{eir.quantum_program_iteration} block symbolically outlines a single quantum circuit iteration, and is enclosed in a for loop expressing 100 iterations. The qubits are reset, a hadamard (H) gate is applied to both, and a controlled-NOT (CX) gate is applied to the qubits with \qinline{\%q1} being the target. The qubit states are read out and stored in the classical bits, and the program result for the current iteration is marked by an \qinline{eir.transmit_results} operation. 

\newpage

\subsection{Two-Qubit GHZ Circuit with Randomized \newline Compilation Ensembling Method}
We apply the randomized compilation error mitigation strategy to the two-qubit GHZ program written in the previous section. The resulting program is depicted graphically in Figure~\ref{fig:comparative}(c). We provide an abbreviated and annotated code snippet for this workload below.

\begin{figure}[h]
    \centering
    \includegraphics[width=1\linewidth]{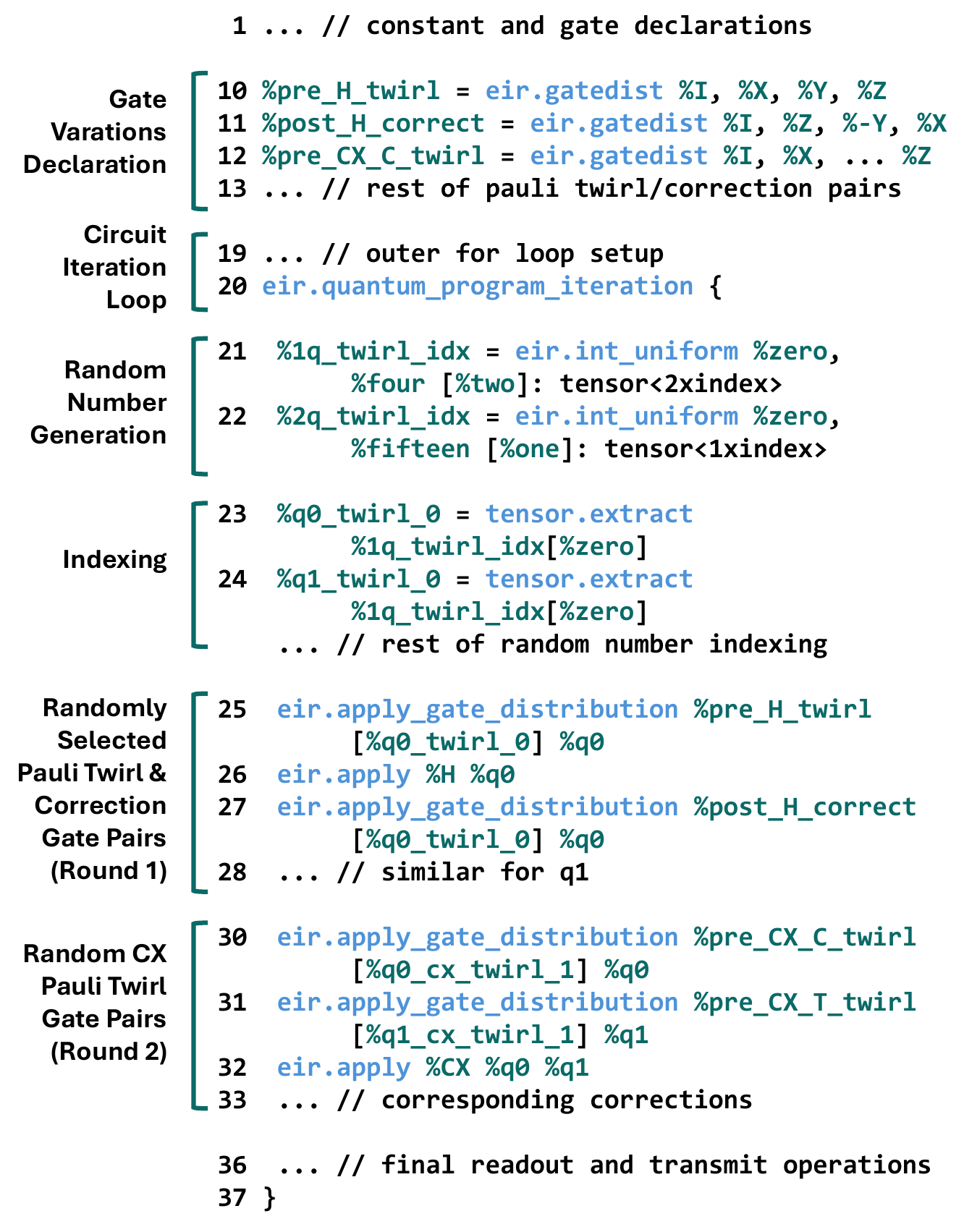}
\end{figure}

The code defines distributions of pauli gates applied before and after each cycle of gates in the original GHZ program. Random numbers are generated and used to select gates to be added in pairs. Single-qubit twirling gates and its respective correction operations are applied before and after each operation in the original GHZ program. Each iteration outlined by the \qinline{eir.quantum_program_iteration} block corresponds to a different realization of the randomized compilation, driven by freshly sampled random numbers.


\end{document}